\documentclass
[prb,twocolumn,showpacs]{revtex4}
\usepackage{graphicx}
\usepackage{amsfonts}
\usepackage{amssymb}

\newcommand{\ba}{\begin{eqnarray}}
\newcommand{\be}{\begin{equation}}
\newcommand{\ea}{\end{eqnarray}}
\newcommand{\ee}{\end{equation}}

\newcommand{\ignore}[1]{}

\begin{document}

\title{Strain-induced large band-gap topological insulator in a new stable silicon allotrope: dumbbell silicene}

\author{Tian Zhang$^{1}$}

\author{Yan Cheng$^{1}$}
\email{ycheng@scu.edu.cn}

\author{Xiang-Rong Chen$^{1}$}
\email{xrchen@scu.edu.cn}

\author{Ling-Cang Cai$^{2}$}

\address{$^{1}$Institute of Atomic and Molecular Physics, College of Physical Science and Technology, Key
             Laboratory of High Energy Density Physics and Technology of Ministry of Education, Sichuan
             University, Chengdu 610065, China\\$^{2}$National Key Laboratory for Shock Wave and Detonation Physics Research, Institute of Fluid Physics, Chinese Academy of Engineering Physics, Mianyang 621900, China}





\date{Oct. 30, 2015}

\begin{abstract}
By the generalized gradient approximation in framewok of density functional theory, we investigate a 2D topological insulator of new silicon allotrope (call dumbbell silicene synthesized recently by Cahangirov $et$ $al$) through tuning external compression strain, and find a topological quantum phase transition from normal to topological insulator, i.e., the dumbbell silicene can turn a two-dimensional topological insulator with an inverted band gap. The obtained maximum topological nontrivial band gap about 12 meV under isotropic strain is much larger than that for previous silicene, and can be further improved to 36 meV by tuning anisotropic strain, which is sufficiently large to realize quantum spin Hall effect even at room-temperature, and thus is beneficial to the fabrication of high-speed spintronics devices. Furthermore, we confirm that the boron nitride sheet is an ideal substrate for the experimental realization of the dumbbell silicene under external strain, maintaining its nontrivial topology. These properties make the two-dimensional dumbbell silicene a good platform to study novel quantum states of matter, showing great potential for future applications in modern silicon-based microelectronics industry.
\end{abstract}


\maketitle
Two-dimensional (2D) materials have been a focus of intense research in recent years \cite{TW1,TW2,TW3,TW4,TW5,TW6,TW7}. As opposed to three-dimensional (3D) one, the optical, electronic, mechanical and thermal properties of the 2D materials are easily adjusted by external strains, defects, electric field, or stacking orders\cite{AJ1,AJ2,AJ3,AJ4}, and thus its realistic performance can be readily improved through current microfabrication technology. More fascinatingly, 2D modes\cite{MODE1,MODE2} was first predicted with quantum spin Hall (QSH) effect, and recently more and more 2D materials have been confirmed as 2D topological insulators (TIs)\cite{TI1,TI2,TI3,TI4,TI5}, also known as QSH insulators. 2D TIs are novel materials characterized by a bulk energy gap and gapless spin-filtered edge states. Different from surface states of 3D TIs, which is only free from exact 180$^{0}$-backscattering and suffers from scattering of other angles, the special edges of 2D TIs are topologically protected by the time reversal symmetry and can immune to nonmagnetic scattering and geometry perturbations, thus open new ways for backscattering-free transport. Clearly, 2D TIs is better than 3D TIs for coherent nondissipative spin transport related applications.

\begin{figure}
\includegraphics[width=3.4in]{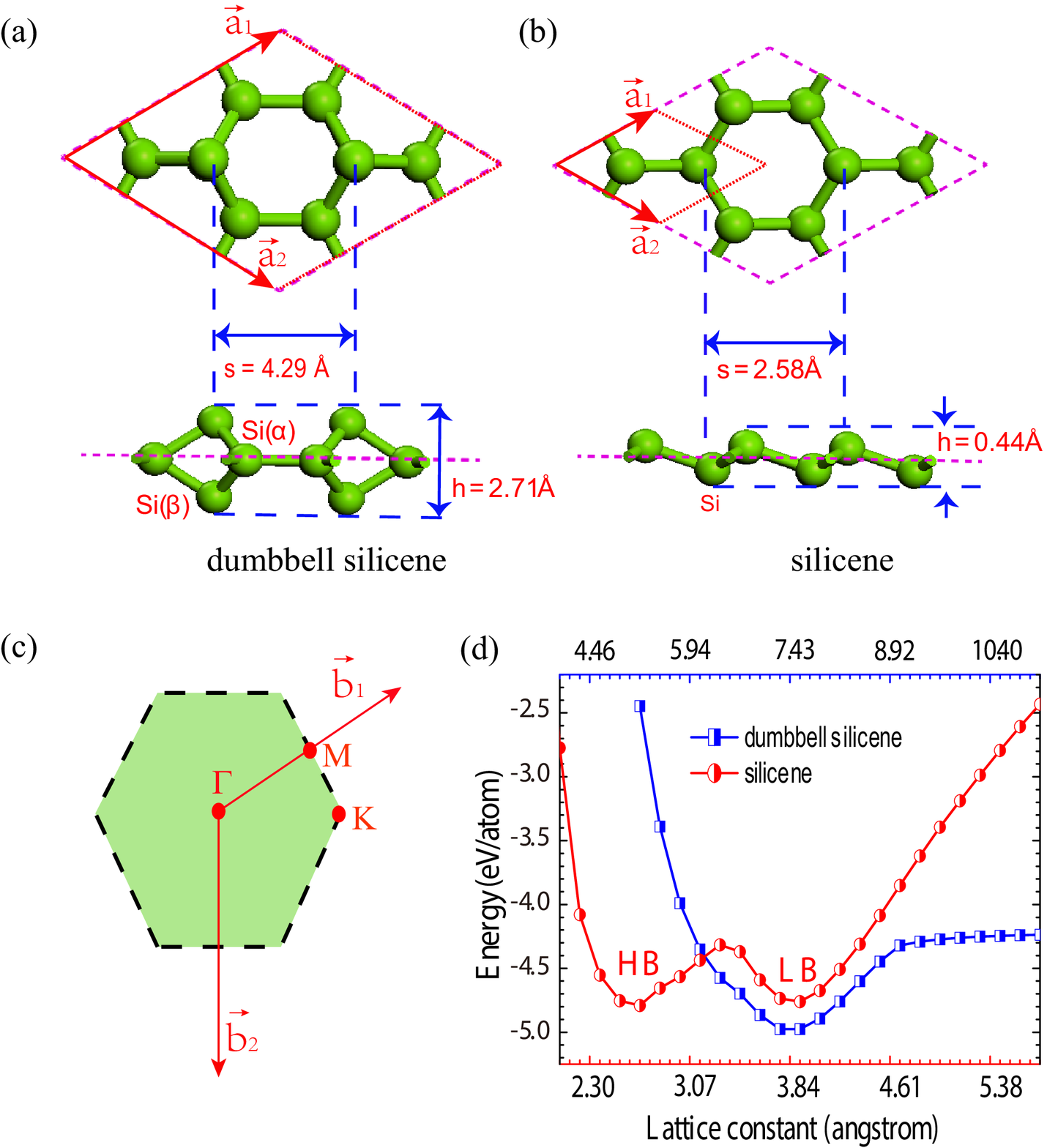}
\caption{(Color online) (a) Top and side view of the dumbbell silicene structure. (b) Top and side view of the silicene structure. Pink dashed lines mark the hexagonal lattice; $\vec{a}_{1}$ and $\vec{a}_{2}$ are the lattice unit vectors. (c) The first Brillouin zone. $\vec{b}_{1}$ and $\vec{b}_{2}$ are the reciprocal lattice unit
 vectors. $\Gamma$, M and K refer to special points in the first Brillouin zone. (d) Total energies of the dumbbell silicene and silicene with different lattice constants.}
\label{fig.sketch}
\end{figure}

The 2D TIs have stimulated enormous research activities in condensed matter physics due to their novel QSH effect and the potential application in quantum computation and spintronics\cite{APP1,APP2}. Well-known 2D TI graphene, a monolayer of carbon atoms forming a similar honeycomb lattice, hosts a miraculous electronic system, and thus becomes perfect breeding ground for a variety of exotic quantum phenomena, such as quantum anomalous Hall effect (QAHE), Majorana fermions and superconductor\cite{QAHE,MJ,SP}. Furthermore, massless Dirac fermions endow graphene with superior carrier mobility\cite{MOB1,MOB2}. Unfortunately, its tiny band gap (about 8$\times$10$^{-4}$ meV\cite{BAND}) opened by spin-orbit coupling (SOC) effect seriously limits its device applications. Subsequently, silicene with a low buckled honeycomb lattice was synthesized and predicted quickly to be a new 2D TI with a relatively large spin-orbit gap of 1.55 meV\cite{SILI1,SILI2}. Almost every striking property of graphene can be transferred to this innovative material. Indeed, these features together with the natural compatibility with current silicon-based microelectronics industry make silicene a promising candidate for future nanoelectronics application. However, its topological nontrivial band gap is still small and limits its room-temperature application in spintronics. Hence, there is great interest in searching for room temperature 2D TI (topological band gap about 26 meV), especially for silicon-based 2D TI.

In this work, we will investigate a 2D topological insulator of new silicon allotrope (call dumbbell silicene synthesized recently by Cahangirov $et$ $al$\cite{DUMB}) through tuning external strain, then find a topological quantum phase transition from normal to topological insulator, accompanied by a band inversion at $\Gamma$ point that changes $Z_{2}$ number from 0 to 1. Isotropic strain can induce the quantum topological phase transition in the dumbbell silicene and modulate its topological nontrivial band gap regularly. The maximum topological nontrivial band gap about 12 meV under isotropic strain can be further improved to 35 meV by tuning anisotropic strain, which is sufficiently large to realize QSH effect of the dumbbell silicene even at room temperature. The pristine dumbbell silicene also owns a Dirac cone at K point originated from $p_{z}$ orbital, like in previous silicene, but the Dirac zone can be easily destroyed as a result of spatial symmetry breaking under anisotropic strain. It implies that the hexagonal symmetry is the main precondition for the present of Dirac cone in the dumbbell silicene. In additional, we further show that the boron nitride (BN) substrate is suitable to support the dumbbell silicen under external strain, and the topologically nontrivial properties of the strained dumbbell silicene can be retained well in the realistic growth on the boron nitride (BN) substrate.

\begin{figure}
\includegraphics[width=2.8in]{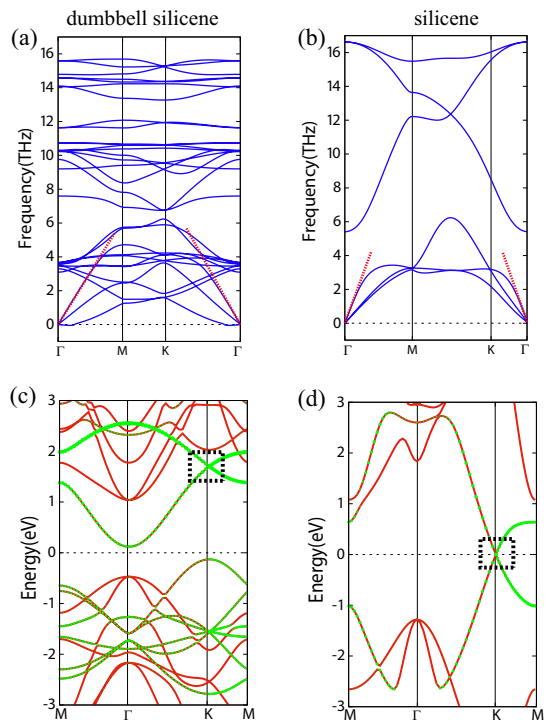}
\caption{(Color online) Vibrational band structure $\omega(\vec{k})$ of (a) the dumbbell silicene and (b) the silicene. The slope of the red dashed lines along the longitudinal acoustic branches near $\Gamma$ corresponds to the speed of sound and the in-plane stiffness. Electronic band structure of (c) the dumbbell silicene and (d) the silicene without SOC. Dirac cone is marked by black rectangular box. The green dots corresponding to Si-$p_{z}$. The Fermi level is set to zero.}
\label{fig.sketch}
\end{figure}

The first-principle calculations based on the density functional theory were performed using the Vienna ab simulation package (VASP)\cite{vasp1,vasp2}. We use the generalized gradient approximations (GGA) of Perdew-Burke-Ernzerhof (PBE)\cite{PBE} for electron-electron interactions and the projector-augmented-wave (PAW)\cite{PAW} pseudo-potentials in the plane-wave basis with an energy cutoff of 700 eV. The Brillouin zone (BZ) was sampled using an 13$\times$13$\times$1 gamma-centered Monkhorst-Pack grid\cite{MK}. A slab model, together with a vacuum layer larger than 30 A, was employed to avoid spurious interactions due to the nonlocal nature of the correlation energy\cite{slab}. The lattice constants are relaxed until the residual force on each atom is less than 0.01 eV/{\AA} and the total energy is less than 1$\times$10$^{-6}$ eV. Spin-orbit coupling (SOC)\cite{SOC} is included in the calculations after the structural relaxations. In additional, we calculated the phonon vibration spectra by using the frozen phonon method as implemented in the PHONOPY code\cite{FP}.

Crystal structure of the dumbbell silicene is presented in Fig. 1(a). Different from silicene [see Fig. 1(b)], in the dumbbell silicene there are two types of Si atoms: one is fourfold coordinated [denoted by Si($\alpha$)], and the other one is only threefold coordinated with a dangling bond [denoted by Si($\beta$)]. It is easy to find that all the Si($\alpha$) atoms keep on the same plane, like in graphene, which is surrounded by the Si($\beta$) atoms to form the dumbbell structure. The optimized dumbbell silicene has a hexagonal structure with space group $D_{6h}$ ($P6/mmm$) and ten Si atoms in one unit cell, which is different from the optimized silicene with space group $D_{3d}$ ($P3m1$) and only two Si atoms in one unit cell. The buckling height of the dumbbell silicene [$h$ in Fig. 1(a) and (b)] about 2.71 {\AA} is much larger than that in the silicene about 0.44 {\AA}, which may result in the larger SOC strength for the dumbbell silicene. The obtained lattice constant and the distance between the neighboring dumbbell units ($s$ in Fig. 1(a)) for the dumbbell silicene are 7.43 {\AA} and 4.29 {\AA}, respectively. These results including those for the silicene are in well agreement with those reported in previous published work\cite{SILI1,DUMB}.

To search for the energetically most stable configuration, we calculate the total energies of the dumbbell silicene and the silicene with different lattice constants [see Fig. 1(c)]. The minimum energy per Si atom for the dumbbell silicene is lower than that for the silicene, implying that the stability of the dumbbell silicene is over the silicene. This may arise mainly from fourfold-coordinated Si atoms because fourfold-coordinated Si is energetically more preferable. Two distinct minima of the silicene in Fig. 1(c) correspond to its low-buckled (LB) and high-buckled (HB) honeycomb structures, where the HB configure is unstable due to its imaginary phonon frequency in a large portion of the Brillouin zone\cite{BHB}. Hence, the mentioned silicene in our whole work refers to the LB silicene. Another way to compare the stability and structural rigidity of the different silicon allotropes is by studying the phonon vibration spectrum. Our results for the phonon vibration spectra of the dumbbell silicene and the silicene are presented in Fig. 2(a) and (b), respectively. The frequencies of all modes are positive over the whole Brillouin zone for the two silicon allotropes, which shows that the two lattice structures are thermodynamically stable and their stability do not depend on the substrates. We compare the slopes of the longitudinal acoustic branches near $\Gamma$, which corresponds to the speed of sound and reveals the in-plane stiffness. As seen in Fig. 2(a), the calculated speed of sound along the $\Gamma$$-$M direction in the dumbbell silicene, $\upsilon_{s}^{\Gamma-M}$=$3.0$ Km/s, is lower than the $\upsilon_{s}^{\Gamma-K}$=$3.3$ Km/s value along the $\Gamma$$-$K direction. The lower rigidity along the $\Gamma$$-$M direction, corresponding to the $\vec{a}_{1}$ + $\vec{a}_{2}$ direction in Fig. 1(a), indicates that compression along this direction requires primarily bond bending, which comes at a lower energy cost than stretching. Different from the dumbbell silicene, the in-plane elastic response of the silicene is nearly isotropic, with nearly the same value $\upsilon_{s}$=2.3 Km/s for the speed of sound along the $\Gamma$$-$M and the $\Gamma$$-$K directions, as shown in Fig. 2(b). We can find that the dumbbell silicene has the larger in-plane stiffness, while the silicene has the higher mechanical flexibility. These results are advantageous when accommodating lattice mismatch during Chemical Vapor Deposition (CVD) growth on a substrate.

\begin{figure}
\includegraphics[width=3.4in]{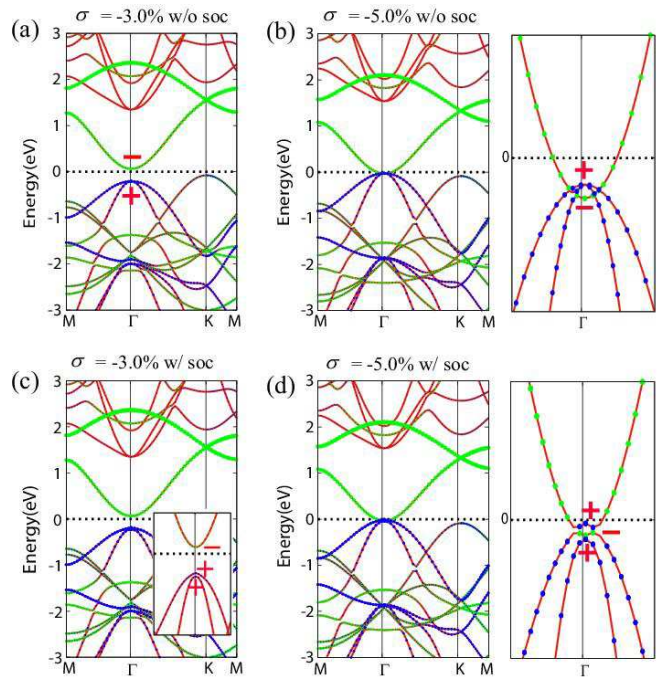}
\caption{(Color online) Electronic band structure of the dumbbell silicene when $\sigma$=$-3.0\%$ (a) without SOC and (c) with SOC, where the inset in (c) shows its enlarged views of electronic band structure around the Fermi level in the vicinity of the $\Gamma$ point. Electronic band structure of the dumbbell silicene when $\sigma$=$-5.0\%$ (b) without SOC and (d) with SOC, where the right-hand panels of (b) and (d) show their enlarged views of electronic band structures around the Fermi level in the vicinity of the $\Gamma$ point. The green and blue dots corresponding to Si-$p_{z}$ and Si-$p_{x,y}$, respectively. (even, odd) parity is denoted by ($+$,$-$) and the Fermi level is set to zero.}
\label{fig.sketch}
\end{figure}

Our DFT results for the electronic band structures without SOC of the two different silicon allotropes are presented in Fig. 2(c) and (d). Band structure information shows that the dumbbell silicene is a semiconductor with an indirect band gap about 0.24 eV, which is much larger than almost zero band gap for the silicene. In additional, it is worth noting that the dumbbell silicene also owns a Dirac cone at K point around the Fermi level, and more surprisingly, like in the silicene, the Dirac cone of the dumbbell silicene also originates from the $p_{z}$ orbital [see the black rectangular box in Fig. 2(c) and (d)]. In fact, the especial Dirac cone is closely related to the Si($\beta$) atoms. Its three of four valence electrons form bonds with neighboring silicon atoms while the remaining one makes a contribution to $p_{z}$ state, which binds covalently with each other and develops into delocalized $\pi$ and $\pi^{*}$ states. Such bands are responsible for Dirac points and linear band dispersion near the Fermi level. However, the Dirac cone of the dumbbell silicene may make less contribution to the transportation, because its other nodes throughout the band structure are parabolic instead of Dirac-type with linear dispersion.

Previous theoretical and experimental studies have shown that the external strain is an excellent method to tune electronic structures of 2D materials. Our results also prove that the band gap of the dumbbell silicene depends sensitively on the applied in-layer isotropic strain $\sigma$, as shown in Fig. 3. The in-plane isotropic strain $\sigma$ is loaded synchronously in the $\vec{a}_{1}$ and $\vec{a}_{2}$ directions, and is defined as $\sigma$ = $(a1/2-a1_{0}/2_{0})/a1_{0}/2_{0}$, where $a1/2$ and $a1_{0}/2_{0}$ are equilibrium lattice constants with and without strain, respectively. Without SOC, the conduction band (CB) and the valence band (VB) at $\Gamma$ point tend to approach together as $\sigma$ increases, and compression beyond 5.0$\%$ shall turn the dumbbell silicene metallic. At $\Gamma$ point, the top of VB is mainly contributed by the antibonding $p_{x\pm iy}$ orbitals with fourfold degeneracy, and the bottom of CB is from the bonding $p_{z}$ orbital. However, by introducing SOC, the fourfold degenerate valence bands are split, as shown in Fig. 3(c). Most remarkably, the closed band gap is reopened when $\sigma$=$-5.0\%$ ($\sigma<0$ means the compression strain), and a band inversion occurs between the $p_{x\pm iy}$ and $p_{z}$ orbitals with a 12 meV nontrivial band gap around $\Gamma$ point, accompanied by the exchange of their parities, as illustrated in Fig. 3(d). The inverted states are labeled by $|p_{z}\rangle$ and $|p_{x\pm iy}\rangle$, and (even, odd) parity is denoted by ($+$,$-$). Such band-inversion character may indicates that the topological phase transition from a trivial state to a nontrivial topological state is happened for the dumbbell silicene as $\sigma$ increases. It is not strange that the Fermi level lies outside the bulk gap for the dumbbell silicene when $\sigma$=$-5.0\%$, because we can artificially adjust the Fermi level inside the gap by applying a gate voltage.

\begin{figure}
\includegraphics[width=3.5in]{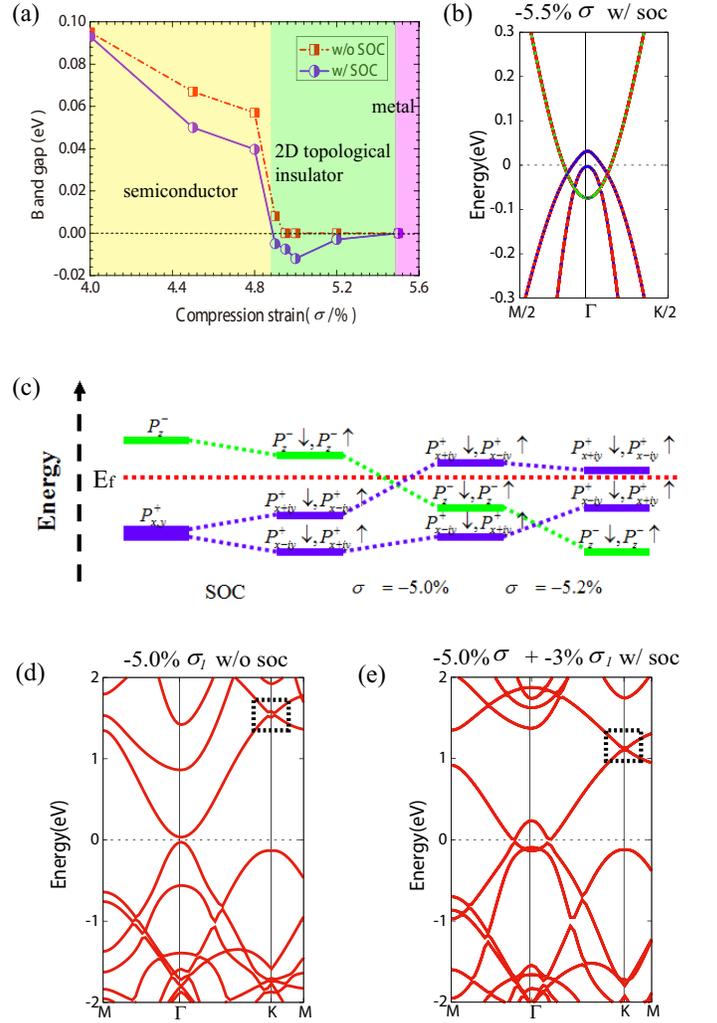}
\caption{(Color online) (a) Dependence of the fundamental band gap for the dumbbell silicene on the in-plane isotropic compression strain $\sigma$. Yellow, green and pink shaded regions indicate the characters of semiconductor, topological insulator and metal for the dumbbell silicene under the isotropic compression strain, respectively. (c) Schematic diagram of the electronic band evolution of the dumbbell silicene with SOC under the isotropic compression strain for the orbitals around the Fermi level at the $\Gamma$ point. Electronic band structure of the dumbbell silicene (b) under $-5.5\%$ $\sigma$ with SOC, (d) under $-3.0\%$ $\sigma_{1}$ without SOC, and (e) under $-5.0\%$ $\sigma$ + $-3.0\%$ $\sigma_{1}$ with SOC. The green and blue dots corresponding to Si-$p_{z}$ and Si-$p_{x,y}$, respectively. The Fermi level is set to zero.}
\label{fig.sketch}
\end{figure}

Further, we apply a rigid method of Fu and Kane\cite{FUKA} to judge whether or not the dumbbell silicene is a 2D topological insulator by calculating its $Z_{2}$ number when $\sigma$ =0 and $-5.0\%$. Such method is valid since the dumbbell silicene has both spatial invention and time reversal symmetries (four time reversal invariant points in the 2D Brillouin zone). Inversion center in the crystal ensures $\varepsilon_{n¦Á}(k)$ =$\varepsilon_{n¦Á}(-k)$, where $\varepsilon_{n¦Á}(k)$ is the electron energy for the $n$-th band with spin index $¦Á$ at $k$ wave vector in the Brillouin zone. The time reversal symmetry makes $\varepsilon_{n¦Á}(k)$ = $\varepsilon_{n\bar{\alpha}}(-k)$, where $\bar{\alpha}$ is the spin opposite to $\alpha$. Hence, we find $\varepsilon_{n¦Á}(k)$ = $\varepsilon_{n\bar{\alpha}}(k)$ when both symmetries are presented, i.e. electronic bands acquire Kramers' doubly degenerate. We state a time-reversal invariant periodic Hamiltonian $H$ with $2N$ occupied bands characterized by Bloch wave functions. A time-reversal operator matrix relates time-reversed wave functions is defined by

\begin{equation}
    A_{\alpha\beta}(\Gamma_{i})=<\mu_{\alpha}(\Gamma_{i})|\Theta|\mu_{\beta}(\Gamma_{-i})>,
\end{equation}
where $\alpha$, $\beta$ =1, 2, ..., $N$, $|\mu_{\alpha}(\Gamma_{i})>$ are cell periodic eigenstates of the
Bloch Hamiltonian, $\Theta$=exp($i\pi$$S_{y})K$ is the time-reversal
operator ($S_{y}$ is spin and $K$ complex conjugation ), which
$\Theta^{2}$=$-$1 for spin 1/2 particles. Since

\begin{equation}
    <\Theta\mu_{\alpha}(\Gamma_{i})|\Theta\mu_{\beta}(\Gamma_{i})>=<\mu_{\beta}(\Gamma_{i})|\mu_{\alpha}(\Gamma_{i})>,
\end{equation}
$A(\Gamma_{i})$ is antisymmetric at TRIM $\Gamma_{i}$. The square of its Pfaffian is equal to its determinant, i.e.,
det[$A$]=Pf$[A]^{2}$. Then $\delta_{i}$=(det$[A(\Gamma_{i})]$)$^{1/2}$/Pf$[A(\Gamma_{i})]$=$\pm$1. Therefore, the topological invariant $Z_{2}$ can be defined as

\begin{equation}
    (-1)^{Z_{2}}=\prod^{M}_{m=1}\xi_{2m}(\Gamma_{i}),
\end{equation}
where $\xi$ is the parities of all occupied bands at $\Gamma_{i}$, and $M$ is the number of Kramers pairs. Results show that the product of parities of occupied bands at the four time reversal invariant points [(0.0, 0.0, 0.0),(0.0, 0.5, 0.0), (0.5, 0.0, 0.0) and (0.5, 0.5, 0.0)] contributes to a +1 parity when $\sigma$=0, yielding a trivial topological invariant $Z_{2}$=0. However, as the strain is increased up to $\sigma$=$-5.0\%$, the product of parities of occupied bands is $-$1 at $\Gamma$ point but +1 at the three other time-reversal invariant momenta. Hence, like in the silicene, the dumbbell silicene also is a 2D topological insulator when the external isotropic compression strain is beyond 5.0$\%$ with $Z_{2}$=1. The evolution of the band gap at $\Gamma$ point with the isotropic compression strain $\sigma$ is presented in Fig. 4(a). We can find that the character of the dumbbell silicene shall undergo a semiconductor, topological insulator to metal process with the increase of external isotropic compression strain $\sigma$. The dumbbell silicene can maintain to be the 2D topological insulator under the isotropic strain $\sigma=-4.9 \sim -5.5\%$. In additional, the dependence of the band gap for the silicene with SOC on the applied isotropic strain $\sigma$ is also analyzed. The band gap of the silicene is not sensitive to $\sigma$. Its Dirac cone at K point is protected by the crystal symmetry, and thus can hardly be eliminated by the isotropic strain $\sigma$. Under the isotropic strain $\sigma=0.0 \sim -6.0\%$, the band gap of the silicene only changes from 1.52 $\sim$ 1.56 meV, which is smaller than the maximum topological band gap about 12 meV of the dumbbell silicene when $\sigma$=$-5.0\%$.  Hence, it is quite promising for the achievement of the QSH effect of the dumbbell silicene at higher temperatures than the silicene.

\begin{figure}
\includegraphics[width=3.4in]{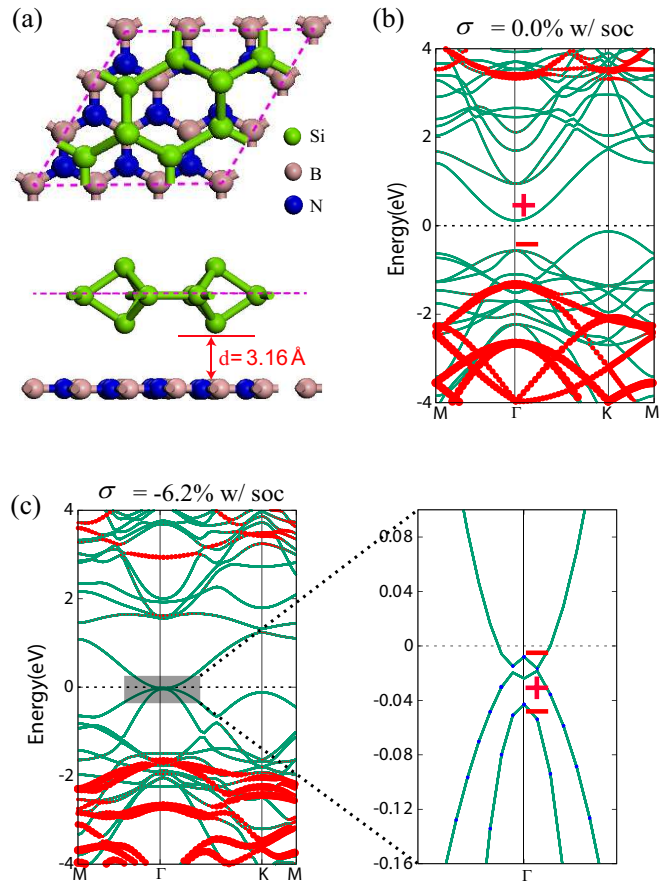}
\caption{(Color online)  (a) Top and side view of the dumbbell silicene structure on the (3$\times$3) BN sheet. Pink dashed lines mark the hexagonal lattice. $d$ is the distance between the dumbbell silicene and substrate. Electronic band structure of the dumbbell silicene on the (3$\times$3) BN sheet with SOC when (b) $\sigma$=0 and (c) $\sigma$=$-6.2\%$. The red and blue dots stand for the contribution from the BN sheet and Si-$p_{x,y}$, respectively. The right-hand panel of (c) show its enlarged view of electronic band structure around the Fermi level in the vicinity of $\Gamma$ point. (even, odd) parity is denoted by ($+$, $-$) and the Fermi level is set to zero.}
\label{fig.sketch}
\end{figure}

We find that the incorporating band inversion in dumbbell silicene could be created solely by external isotropic strain, even without considering the SOC, as shown in Fig. 3(b). As the isotropic strain $\sigma$ is increased up to $-5.0\%$, the top of VB ($|p_{x\pm iy}\rangle$ state) and the bottom of CB ($|p_{z}\rangle$ state) is inversion, and then the exchanged bottom of VB ($|p_{z}\rangle$ state) and the previous degenerate top of VB ($|p_{x\mp iy}\rangle$ state) is rapidly once again inversion due to the existence of the fourfold degenerate valence bands at $\Gamma$ point, which turns the dumbbell silicene metallic. Hence, it is impossible that the dumbbell silicene can turn 2D topological insulator solely by external isotropic strain. SOC plays a vital role in lifting the band degeneracy for the valence bands at $\Gamma$ point, and thus creating a gap at the crossing points originating from the band inversion. However, the SOC in Si is too weak, which leads to a tiny energy difference between the $|p_{x\pm iy}\rangle$ and $|p_{x\mp iy}\rangle$ states. As seen in Fig. 4(b) and (c), the exchanged $|p_{z}\rangle$ state and the $|p_{x\mp iy}\rangle$ state tend to approach together as the compression strain $\sigma$ is increased, then exchange secondly their orbital compositions, similar to the case without SOC. Hence, we know that the weak SOC strength and the fourfold degenerate valence bands at $\Gamma$ point are the main restricted factors for the dumbbell silicene to create larger topological band gap.

One restricted factor for the fourfold degenerate valence bands of the dumbbell silicene can be solved by tuning anisotropic strain. The calculated results for the dumbbell silicene under anisotropic strain $\sigma_{x}$ are presented in Fig. 4(d) and (e). The anisotropic strain $\sigma_{1}$ is only along the $\vec{a}_{1}$ direction and is defined as $\sigma_{1}$ = $(a1-a1_{0})/a1_{0}$, where $a1$ and $a1_{0}$ are equilibrium lattice constants along the $\vec{a}_{1}$ direction with and without strain, respectively. Results show that the fourfold degenerate valence band at $\Gamma$ point for the dumbbell silicene can be lifted solely by anisotropic strain $\sigma_{1}$, as shown in Fig. 4(d). It may create the larger topological band gap than that only induced by $\sigma$. Hence, we apply an extra anisotropic strain $\sigma_{1}$ about $-3.0\%$ on the previous lattice structure under $\sigma=-5.0\%$ with the largest nontrivial topological band gap, as shown in Fig. 4(e). The optimized lattice structure of the dumbbell silicene under $-5.0\%$ $\sigma$ and $-3.0\%$ $\sigma_{1}$ still has both spatial invention and time reversal symmetries, and hence its topological property still can be judged by the previous method of Fu and Kane. We find that the character of 2D topological insulator for the dumbbell silicene is retained and the nontrivial topological band gap is increased up to 36 meV, which is sufficiently larger to realize QSH effect at room-temperature. Unfortunately, with the further increase of the anisotropic compression strain $\sigma_{1}$, the nontrivial topological band gap for the dumbbell silicene under $\sigma=-5.0\%$ is decreased. In additional, it is worth to noting that the Dirac cone for the dumbbell silicene under the anisotropic strain $\sigma_{1}$ are destroyed as a result of spatial symmetry breaking, different from the previous case under the isotropic strain $\sigma$, as shown in Fig. 4(e). It is strong verified that hexagonal symmetry is the main precondition for the present of Dirac cone in the dumbbell silicene.

The in-plane strain on the dumbbell silicene can be realized by bending its flexure substrate in experiment, similar to graphene\cite{STRAIN}, where the amount of the strain is proportional to 2D mode position of the dumbbell silicene. The coupling between the substrate and the topological insulators deposited on them may destroy the topological nontrivially or increase the SOC band gap. Hence, it is important to find a proper substrate for the dumbbell silicene on which its exotic topological properties under high external strain can be retained for future applications. Because the BN sheet has a close lattice structure and large band gap and high dielectric constant, here we use it as a substrate to support DB stanene. Fig. 5(a) shows the geometrical structure of the dumbbell silicene on the (3$\times$3) BN sheet, where the lattice mismatch is less than 1.5$\%$. The dumbbell silicene almost retains the original structure with lattice constant about 7.50 {\AA} and a buckling height $h$ about 2.69{\AA}. The
distance between the adjacent layers [$d$ in Fig. 3(b)] is about 3.16 {\AA}. Cohesive energy for the heterostructure of the dumbbell silicene and the hexagonal BN sheet is defined by $E_{coh}$=$E_{total}$-$E_{DS}$-$E_{BN}$, where $E_{DS}$ and $E_{BN}$ are the total energy of single dumbbell silicene and single hexagonal BN sheet, respectively. The obtained cohesive energy is only about $-$0.58 eV per atom, showing a weak interaction between the dumbbell silicene and the BN sheet. The calculated band structure of the heterostructure with SOC under $\sigma$=0 is shown in Fig. 5(b). We find that the dumbbell silicene on BN sheet remains semiconducting. There is essentially no charge transfer between the adjacent layers, and the states around the Fermi level are dominantly contributed by the dumbbell silicene. As the isotropic strain is increased up to $-$6.2$\%$,  the top of VB and the bottom of CB for the dumbbell silicene on BN sheet are inversion, and thus open a small topological band gap [see the inset of Fig. 5(c)],  accompanied by the exchange of their parities. Evidently, the dumbbell silicene on the (3$\times$3) BN substrate also is a 2D topological insulator whose band inversion is not affected by the substrate. Although the required isotropic compression strain to realize the quantum topological phase transition is increased for the dumbbell silicene on BN sheet due to the inevitable lattice mismatch, we still think that the hexagonal BN sheet is a suitable substrate to support the dumbbell silicene, maintaining its nontrivial topology under high compression strain.

In summary, based on first-principles calculations, we have proven that the dumbbell silicene is an excellent 2D topological material when the isotropic compression strain $\sigma$ loaded synchronously in the $\vec{a}_{1}$ and $\vec{a}_{2}$ directions is larger than about $5.0\%$. The pristine structure of the dumbbell silicene is found to be an indirect band gap semiconductor and is more mechanical stable than the previous silicene. Its electronic band can be inverted between the valence band maximum of $p_{x,y}$ orbitals and the conduction band maximum of $p_{z}$ orbital at $\Gamma$ point by tuning isotropic compression strain $\sigma$, accompanied by the change of $Z_{2}$ number from 0 to 1, and thus leads to quantum spin Hall (QSH) state of the dumbbell silicene. The obtained maximum topological nontrivial band gap about 12 meV under isotropic compression strain $\sigma$ can be further improved to 35 meV by tuning anisotropic strain $\sigma_{1}$, which is useful for the application of QSH effect at room temperature and the fabrication of high-speed spintronics devices. A Dirac cone at K point originated from $p_{z}$ orbital is found in the pristine dumbbell silicene, similar to previous silicene, but can be easily destroyed as a result of spatial symmetry breaking under anisotropic strain. It implies that the hexagonal symmetry is the main precondition for the present of Dirac cone in the dumbbell silicene. In additional, we confirm that the boron nitride (BN) substrate is suitable to support the dumbbell silicen under external strain, and the topologically nontrivial properties of the strained dumbbell silicene are retained well in the realistic growth on the boron nitride (BN) substrate. We expect that the previous silicene can be replaced by the dumbbell silicene in modern silicon-based microelectronics industry, and the quantum anomalous Hall effect, Chern half metallicity, and topological superconductivity can also be realized in the dumbbell silicene, which shall make supported dumbbell silicene an ideal platform to study quantum states of matter and show great potential for future applications.

The authors would like to thank the support by the NSAF Joint Fund Jointly set up by the National Natural Science Foundation of China and the Chinese Academy of Engineering Physics (Grant Nos. U1430117, U1230201). Some calculations are performed on the ScGrid of Supercomputing Center, Computer Network Information Center of Chinese Academy of Sciences.


\begin{thebibliography}{99}

\bibitem{TW1} J. Guan, Z. Zhu, and D. Tom\'{a}nek, Phys.\
Rev.\ Lett.\ \textbf{113}, 046804 (2014).

\bibitem{TW2} F. H. L. Koppens \textit{et al}., Nat.\
Nanotechnol.\ \textbf{9}, 780 (2014).

\bibitem{TW3} D. Jariwala \textit{et al}., ACS\ Nano\ \textbf{8}, 1102 (2014).

\bibitem{TW4} Y. -C. Lin, D. O. Dumcenco, Y. -S. Huang, and K. Suenaga, Nat.\
Nanotechnol.\ \textbf{9}, 391 (2014).

\bibitem{TW5} L. Li \textit{et al}., Nat.\
Nanotechnol.\ \textbf{9}, 372 (2014).

\bibitem{TW6} S. Banerjee, G. Periyasamy, and S. K. Pati, J.\ Mater.\ Chem.\ A\ \textbf{2}, 3856 (2014).

\bibitem{TW7} P. Y. Huang \textit{et al}., Nano\ Lett.\ \textbf{12}, 1081 (2012).

\bibitem{AJ1} S. Zhao, W. Kang, and J. Xue, Appl.\ Phys.\ Lett.\ \textbf{104}, 113106 (2014).

\bibitem{AJ2} C. Ataca, H. Sahin, E. Akt\"{u}rk, and S. Ciraci, J.\ Phys.\ Chem.\ C\ \textbf{115}, 3934 (2011).

\bibitem{AJ3} Eduardo. V. Castro \textit{et al}., Phys.\
Rev.\ Lett.\ \textbf{99}, 216802 (2007).

\bibitem{AJ4} G. Constantinescu, A. Kuc, and T. Heine, Phys.\
Rev.\ Lett.\ \textbf{111}, 036104 (2013).

\bibitem{MODE1} C. L. Kane, and E. J. Mele, Phys.\ Rev.\ Lett.\ \textbf{95}, 226801 (2005).

\bibitem{MODE2} B. A. Bernevig, T. L. Hughes, and S. -C. Zhang, Science\ \textbf{314}, 1757 (2006).

\bibitem{TI1} T. Zhang, J. -H. Lin, Y. -M. Yu, X. -R. Chen, and W. -M. Liu, Scientific\ Reports\ \textbf{5}, 13927 (2015).

\bibitem{TI2} W. Luo, and H. Xiang, Nano\ Lett.\ \textbf{15}, 3230 (2015).

\bibitem{TI3} F. -C. Chuang \textit{et al}., Nano\ Lett.\ \textbf{14}, 2505 (2014).

\bibitem{TI4} X. Qian, J. Liu, L. Fu, and J. Li, Science\ \textbf{346}, 1344 (2014).

\bibitem{TI5} C. Niu, G. Bihlmayer, H. Zhang, D. Wortmann, S. Blugel, and Y. Mokrousov, Phys.\ Rev.\ B\ \textbf{91},041303 (2015).

\bibitem{APP1} M. Z. Hasan, and C. L. Kane, Rev.\ Mod.\ Phys.\
\textbf{82}, 3045 (2010).

\bibitem{APP2} X. -L. Qi, S. -C. Zhang, and C. L. Kane, Rev.\ Mod.\ Phys.\
\textbf{83}, 1057 (2011).

\bibitem{QAHE} J. Zhang, B. Zhao, Y. Yao, and Z. Yang, Scientific\ Reports\
\textbf{5}, 10629 (2015).

\bibitem{MJ} A. M. Black-Schaffer, Nature\
\textbf{109}, 197001 (2012).

\bibitem{SP} R. Nandkishore, L. S. Levitov, and A. V. Chubukov, Nat.\ Phys.\
\textbf{8}, 2208 (2012).

\bibitem{MOB1} Y. Wu, \textit{et al}., Nat.\ Phys.\
\textbf{472}, 74 (2011).

\bibitem{MOB2} K. F. Mak, C. Lee, J.  Hone, J. Shan, and T. F.  Heinz, Phys.\ Rev.\
Lett.\ \textbf{105},  136805 (2010).

\bibitem{BAND} Y. Yao, F. Ye, X. Qi, S. Zhang, and Z. Fang, Phys.\
Rev.\ B\ \textbf{75}, 041401 (2007).

\bibitem{SILI1} X. -L. Zhang, L. -F. Liu, and W. -M. Liu, Scientific\ Reports\ \textbf{3}, 02908 (2013).

\bibitem{SILI2} C. Liu, W. Feng, and Y. Yao, Phys.\
Rev.\ B\ \textbf{107}, 076802 (2011).

\bibitem{DUMB} S. Cahangirov \textit{et al}., Phys.\
Rev.\ B \ \textbf{90}, 035448 (2014).

\bibitem{vasp1} G. Kresse, and J. Furthmuller, Phys.\
Rev.\ B\ \textbf{54}, 11169 (1996).

\bibitem{vasp2}G. Kresse, and J. Furthmuller, Comput.\ Mater.\ Sci.\ \textbf{6}, 15 (1996).

\bibitem{PBE}J. P. Perdew, K. Burke, and M. Ernzerhof, Phys.\ Rev.\
Lett.\ \textbf{77}, 3865 (1996).

\bibitem{PAW}P. E. Blochl, Phys.\ Rev.\ B\ \textbf{50},
17953 (1994).

\bibitem{MK}H. J. Monkhorst, and J. D. Pack, Phys.\ Rev.\
B\ \textbf{13}, 5188 (1976).

\bibitem{slab}T. Olsen, J. Yan, J. J. Mortensen, and K. S. Thygesen, Phys.\ Rev.
\ Lett.\ \textbf{107}, 156401 (2011).

\bibitem{SOC}A. Togo, F. Oba, and I. Tanaka, Phys.\ Rev.\ B \textbf{78}, 134106
(2008).

\bibitem{FP}K. Parlinski, Z. Q. Li, and Y. Kawazoe, Phys.\ Rev.\ Lett.\ \textbf{78}, 4063 (1997).

\bibitem{BHB}S. Cahangirov, M. Topsakal, E. Akt\"{u}rk, H. Sahin, and S. Ciraci, Phys.\ Rev.\ Lett.\ \textbf{102}, 236804 (2009).

\bibitem{FUKA}L. Fu, and C. L. Kane, Phys.\ Rev.\ B \textbf{76}, 045302
(2007).

\bibitem{STRAIN}T. Yu, Z. Ni, C. Du, Y. You, Y. Wang, and Z. Shen, J.\ Phys.\ Chem.\ Lett.\ \textbf{112}, 12602
(2008).

\end{thebibliography}
\end{document}